        \documentstyle[twocolumn,amsfonts,euscript]{article}
        \newcommand{\al}{\alpha}
        \newcommand{\gam}{\gamma}
        \newcommand{\del}{\delta}
        \newcommand{\eps}{\epsilon}
        \newcommand{\veps}{\varepsilon}
        \newcommand{\lam}{\lambda}
        \newcommand{\vph}{\varphi}
        \newcommand{\sig}{\sigma}
        \newcommand{\vpi}{\varpi}
        \newcommand{\th}{\theta}
        \newcommand{\vth}{\vartheta}
        \newcommand{\ups}{\upsilon}
        \newcommand{\om}{\omega}
        \newcommand{\Del}{{\mathit \Delta}}
        \newcommand{\Gam}{{\mathit \Gamma}}
        \newcommand{\Lam}{{\mathit \Lambda}}
        \newcommand{\Ph}{{\mathit \Phi}}
        \newcommand{\Sig}{{\mathit \Sigma}}
        \newcommand{\Th}{{\mathit \Theta}}
        \newcommand{\XI}{{\mathit \Xi}}
        \newcommand{\Ups}{{\mathit \Upsilon}}
        \newcommand{\Om}{{\mathit \Omega}}

        \newcommand{\RE}{{\mathbb{R}}}
        \newcommand{\ZZ}{{\mathbb{Z}}}
        \newcommand{\g}{{\mathfrak g}}
        \newcommand{\s}{s}
        \newcommand{\A}{{\EuScript{A}}}
        \newcommand{\DD}{{\EuScript{D}}}
        \newcommand{\EE}{{\EuScript{E}}}
        \newcommand{\FF}{{\EuScript{F}}}
        \newcommand{\GG}{{\EuScript{G}}}
        
        \newcommand{\MM}{{\EuScript{M}}}
        \newcommand{\OO}{{\EuScript{O}}}
        \newcommand{\SS}{{\EuScript{S}}}

        \newcounter{sect}
        \newcounter{subsect}
        \newcommand{\sect}[1]{\vspace{2ex}
                \addtocounter{sect}{1}\setcounter{subsect}{0}
                \begin{flushleft}
                {{\large\bf \arabic{sect}. {#1}}}
                \end{flushleft}
                \setcounter{equation}{0}
                \def\theequation{\arabic{sect}.\arabic{equation}}
                \def\thefigure{\arabic{sect}.\arabic{figure}}}
        \newcommand{\subsect}[1]{\vspace{0.25ex}\addtocounter{subsect}{1}
                \begin{flushleft}
                {{\bf \arabic{sect}.\arabic{subsect} {#1}}}
                \end{flushleft}}
        \newcommand{\be}{\begin{equation}}
        \newcommand{\ee}{\end{equation}}
        \newcommand{\bea}{\begin{eqnarray}}
        \newcommand{\eea}{\end{eqnarray}}
        \newcommand{\nno}{\nonumber \\}
        \newcommand{\sep}[1]{\!\!\!\! &{#1}& \!\!\!\! }
        \newcommand{\eq}{\sep{=}}
        \newcommand{\vc}{\sep{ }}

        \newcommand{\bra}{\langle}
        \newcommand{\ket}{\rangle}
        \newcommand{\m}{\backslash}
        \newcommand{\gtimes}{\hat\otimes}

        \newcommand{\medwedge}{\mbox{\fontsize{12pt}{0pt}\selectfont $\wedge$}}
        \newcommand{\hf}{{\textstyle \frac{1}{2}}}
        \newcommand{\qt}{{\textstyle \frac{1}{4}}}
	\newcommand{\hfi}{{\textstyle \frac{\ii}{2}}}
        
        \newcommand{\ii}{\sqrt{-1}}
        \newcommand{\ch}{{\,\mathrm{ch}\,}}
        \newcommand{\ev}{{\mathrm{ev}}}
        \newcommand{\grad}{{\mathrm{grad}\,}}
        \newcommand{\tr}{{\,\mathrm{tr}}}
        \newcommand{\str}{{\,\mathrm{str}}}
        \newcommand{\vol}{{\mathrm{vol}}}
        \newcommand{\cl}{{\mathrm{Cl}}\,}
        
        \newcommand{\pdr}{\partial}
        \newcommand{\bos}{|_{\th=0}}
        
        \newcommand{\bfx}{{\mbox{\boldmath $x$}}}
        \newcommand{\bfg}{{\mbox{\boldmath $\gam$}}}
        \newcommand{\sbfx}{{\mbox{\boldmath ${\scriptstyle x}$}}}
        \newcommand{\sbfg}{{\mbox{\boldmath ${\scriptstyle \gam}$}}}
        \newcommand{\Ad}{{\mathrm{Ad}}}
        \newcommand{\Pf}{{\mathrm{Pf}}}
        \newcommand{\End}{{\mathrm{End}}}
        \newcommand{\Map}{{\mathrm{Map}}}
        \newcommand{\Lie}{{\mathrm{Lie}}}
        
        \newcommand{\MQ}{{\mathrm{MQ}}}
        \newcommand{\AJ}{{\mathrm{AJ}}}
        \newcommand{\CMR}{{\mathrm{CMR}}}
        
        \newcommand{\bint}{\int^B\!\!}
	\newcommand{\gh}{{\hat g}}
        \newcommand{\RF}{{\RE^{0|1}}}
        \newcommand{\two}[4]{\left\{    \begin{array}{ll}
                                        {#1}, & {\mbox{if }} {#2}, \\
                                        {#3}, & {\mbox{if }} {#4}
                                        \end{array}     \right.}

\begin{document}

        \begin{center}
{\LARGE\sf MATHAI-QUILLEN FORMALISM}\\

        \vspace{4ex}
        {\large\rm Siye Wu}

        {\em Department of Mathematics, University of Colorado, Boulder, 
          CO 80309-0395, USA}\\

        \end{center}

\sect{Introduction}

Characteristic classes play an essential role in the study of global 
properties of vector bundles.
Particularly important is the Euler class of real orientable vector bundles. 
A de Rham representative of the Euler class (for tangent bundles) first
appeared in Chern's generalisation of the Gauss-Bonnet theorem to higher
dimensions.
The representative is the Pfaffian of the curvature, whose cohomology class
does not depend on the choice of connections.
The Euler class of a vector bundle is also the obstruction to the existence
of a no-where vanishing section.
In fact, it is the Poincar\'e dual of the zero-set of any section which
intersects the zero section transversely.
In the case of tangent bundles, it counts (algebraically) the zeros of a
vector field on the manifold.
That this is equal to the Euler characteristic number is known as the Hopf
theorem.
Also significant is the Thom class of a vector bundle: it is the Poincar\'e
dual of the zero section in the total space.
It induces, by a cup product, the Thom isomorphism between the cohomology of 
the base space and that of the total space with compact vertical support.
Thom isomorphism also exists and plays an important role in K-theory.

Mathai and Quillen (1986) obtained a representative of the Thom class
by a differential form on the total space of a vector bundle.
Instead of having a compact support, the form has a nice Gaussian peak near
the zero section and exponentially decays along the fiber directions.
The pull-back of Mathai-Quillen's Thom form by any section is 
a representative of the Euler class.
By scaling the section, one obtains an interpolation between the Pfaffian
of the curvature, which distributes smoothly on the manifold, and the
Poincar\'e dual of the zero-set, which localises on the latter.
This elegant construction prove to be extremely useful in many situations,
from the study of Morse theory, analytic torsion in mathematics to the
understanding of topological (cohomological) field theories in physics.

In this article, we begin with the construction of Mathai-Quillen's Thom form. 
We also consider the case with group actions, with a review of equivariant 
cohomology and then Mathai-Quillen's construction in this setting.
Next, we show that much of the above can be formulated as a ``field theory''
on a superspace of one fermionic dimension.
Finally, we present the interpretation of topological field theories using
the Mathai-Quillen formalism.

\sect{Mathai-Quillen's Construction} 

\subsect{Berezin integral and supertrace}

Let $V$ be an oriented real vector space of dimension $n$ with a volume 
element $\nu\in\medwedge^nV$ compatible with the orientation.
The {\em Berezin integral} of a form $\om\in\medwedge^*V^*$ on $V$,
denoted by $\int^B\om$, is the pairing $\bra\nu,\om\ket$.
Clearly, only the top degree component of $\om$ contributes. 
For example, if $\sig\in\medwedge^2V^*$ is a $2$-form, then 
\[ \bint e^\sig=\two{\bra\nu,\frac{\sig^{\wedge(n/2)}}{(n/2)!}\ket}
   {{\mbox{$n$ is even}}}{0}{{\mbox{$n$ is odd.}}} \]
If $V$ has a Euclidean metric $(\cdot,\cdot)$, then $\nu$ is chosen to be
of unit norm.
If $\Sig\in\End(V)$ is skew-symmetric, then $\hf(\,\cdot\,,\Sig\,\cdot\,)$
is a $2$-form and, if $n$ is even, the {\em Pfaffian} of $\Sig$ is 
$$\Pf(\Sig)=\bint e^{\frac{1}{2}(\,\cdot\,,\Sig\,\cdot\,)}.$$

The Berezin integral can be defined on elements in a graded tensor product 
$\medwedge^*V^*\gtimes A$, where $A$ is any $\ZZ_2$-graded commutative
algebra. 
For example, if we consider the identity operator $\bfx=\mathrm{id}_V$ as
a $V$-valued function on $V$, then $d\bfx$ is a $1$-form on $V$ valued
in $V$, and $(d\bfx,\cdot\,)$ is a $1$-form valued in $V^*$.
Let $\{e_1,\dots,e_n\}$ be an orthonormal basis of $V$ and write
$\bfx=x^ie_i$, where $x^i$ are the coordinate functions on $V$.
We let
$$u(\bfx)=\frac{(-1)^{\frac{n(n+1)}{2}}}{(2\pi)^{\frac{n}{2}}}
  \bint e^{-\frac{1}{2}(\sbfx,\sbfx)-(d\sbfx,\,\cdot\,)}. $$
The integrand is in $\Om^*(V)\gtimes\medwedge^*V^*$.
The result is
\be\label{gaussian}
u(\bfx)=\frac{1}{(2\pi)^{\frac{n}{2}}}e^{-\frac{1}{2}(\sbfx,\sbfx)}
   \,dx^1\wedge\cdots\wedge dx^n,
\ee
a Gaussian $n$-form whose (usual) integration on $V$ is $1$.

Let $\cl(V)$ be the Clifford algebra of $V$.
For any orthonormal basis $\{e_i\}$, let $\gam^i$ be the corresponding
generators of $\cl(V)$ and let $\bfg=e_i\otimes\gam^i\in V\otimes\cl(V)$.
For any $\om\in\medwedge^kV^*$, we have $\om(\bfg,\dots,\bfg)=
\frac{1}{k!}\om_{i_1\cdots i_k}\gam^{i_1}\cdots\gam^{i_k}\in\cl(V)$.
If $n$ is even, the Clifford algebra has a unique $\ZZ_2$-graded
irreducible spinor representation $S(V)=S^+(V)\oplus S^-(V)$.
For any element $a\in\cl(V)$, the {\em supertrace} is
$\str\,a=\tr_{S^+(V)}\,a-\tr_{S^-(V)}\,a$.
If $\Sig\in\End(V)$ is skew-symmetric, then
$$ \str\;e^{\frac{\ii}{4}(\sbfg,\Sig\sbfg)}=
\hat A(\Sig)^{-\frac{1}{2}}\,\Pf(\Sig), $$
where $\hat A(\Sig)=\det\left(\frac{\Sig/2}{\sinh(\Sig/2)}\right)$.
More generally, supertrace can be defined on $\cl(V)\gtimes A$
for any $\ZZ_2$-graded commutative algebra $A=A^+\oplus A^-$.
If $\Sig$ is skew-symmetric and $\al\in V^*\otimes A^-$, then
\be\label{supertrace}
\str\;e^{\frac{\ii}{4}(\sbfg,\Sig\sbfg)+(\frac{\ii}{2})^{\frac{1}{2}}
\al(\sbfg)}=
\hat A(\Sig)^{-\frac{1}{2}}\!\bint e^{\frac{1}{2}(\cdot,\Sig\cdot)+\al}.
\ee

\subsect{Representatives of the Euler and Thom classes}

Let $M$ be a smooth manifold and let $\pi\colon E\to M$ be an oriented 
real vector bundle of rank $r$. 
Suppose $E$ has a Euclidean structure $(\cdot,\cdot)$ and $\nabla$ is 
a compatible connection.
The curvature $R\in\Om^2(M,\End(E))$ is skew-symmetric, and hence
$(\,\cdot\,,R\,\cdot\,)\in\Om^2(M,\medwedge^2E^*)$.
A de Rham representative of the Euler class of $E$ is
\be\label{euler}
e_\nabla(E)=\frac{1}{(2\pi)^\frac{r}{2}}\bint
e^{\frac{1}{2}(\,\cdot\,,R\,\cdot\,)}
=\Pf\left({\textstyle \frac{R}{2\pi}}\right).
\ee
Here the Berezin integration is fiberwise in $E$: it is the pairing 
between the integrand and the unit section $\nu$ of the trivial line
bundle $\medwedge^rE$ that is consistent with the orientation of $E$.
The de Rham cohomology class of (\ref{euler}) is independent of the 
choice of $(\cdot,\cdot)$ or $\nabla$.

Let $s$ be a section of $E$.
Following Berline, Getzler and Vergne (1992) and Zhang (2001), we consider 
\be\label{mqexp}
\SS_{\nabla\!,s}=\hf(s,s)+(\nabla s,\cdot\,)+\hf(\,\cdot\,,R\,\cdot\,),
\ee
a differential form on $M$ valued in $\medwedge^*E^*$.
Mathai-Quillen's representative of the Euler class is
\be\label{mq-euler}
e_{\nabla\!,s}(E)=\frac{(-1)^{\frac{r(r+1)}{2}}}{(2\pi)^{\frac{r}{2}}}
                \bint e^{-\SS_{\nabla\!,s}}.
\ee
One can show that $e_{\nabla\!,s}(E)$ is closed and that as $\beta$ varies,
the cohomology class of $e_{\nabla\!,\beta s}(E)$ does not change.
By taking $\beta\to0$, the de Rham class of $e_{\nabla\!,s}(E)$ 
is equal to that of $e_\nabla(E)$ when $r$ is even.
The form $e_{\nabla\!,\beta s}(E)$ provides an continuous interpolation 
between (\ref{euler}) and the limit as $\beta\to\infty$, when the form 
is concentrated on the zero locus of the section $s$.
In fact, the Euler class is the Poincar\'e dual to the homology class
represented by $s^{-1}(0)$.
Hence if $n\ge m$ and if $\om\in\Om^{n-m}(M)$ is closed, we have
\be\label{mq-pd}
\int_M\om\wedge e_{\nabla\!,s}(E)=\int_{s^{-1}(0)}\om
\ee
when $s$ intersects the zero section transversely.

To obtain Mathai-Quillen's representative of the Thom class, we consider
the pull-back of $E$ to $E$ itself.
The bundle $\pi^*E\to E$ has a tautological section $\bfx$.
Applying the (\ref{mq-euler}) to this setting, we get
\be\label{mq-thom}
\tau_\nabla(E)=\frac{(-1)^{\frac{r(r+1)}{2}}}{(2\pi)^{\frac{r}{2}}} 
    \bint e^{-\frac{1}{2}(\sbfx,\sbfx)-(\nabla\sbfx,\cdot\,)
                  -\frac{1}{2}(\,\cdot\,,R\,\cdot\,)},
\ee
where $(\cdot,\cdot)$, $\nabla$ and $R$ are understood to be the pull-backs
to $\pi^*E$.
This is a closed form on the total space of $E$.
Moreover, its restriction to each fiber is the Gaussian form (\ref{gaussian}).
The cohomology groups of differential forms with exponential decay 
along the fibers are isomorphic to those with compact vertical support
or the relative cohomology groups $H^*(E,E\m M)$.
Here $M$ is identified with its image under the inclusion $i\colon M\to E$
by the zero section. 
Under the above isomorphism, the cohomology class represented by 
$\tau_\nabla(E)$ coincides with the Thom class $\tau(E)=i_*1\in H^r(E,E\m M)$
defined topologically.
For any section $s\in\Gam(E)$, we have 
$e_{\nabla\!,s}(E)=s^*\tau_\nabla(E)$.

\subsect{Character form of the Thom class in K-theory}

Let $E=E^+\oplus E^-$ be a $\ZZ_2$-graded vector bundle over $M$.
The spaces $\Om^*(M,E)$, $\Gam(\End(E))$ and $\Om^*(M)\gtimes\Gam(\End(E))$
are also $\ZZ_2$-graded.
The action of $\al\gtimes T\in\Om^*(M)\gtimes\Gam(\End(E))$ on 
$\beta\otimes s\in\Om^*(M,E)$ is
$$ \al\gtimes T\,\colon\,\beta\otimes s\mapsto(-1)^{|T|\,|\beta|}
(\al\wedge\beta)\otimes(Ts). $$
The {\em supertrace} of $A\in\Gam(\End(E))$ is 
$\str\,A=\tr_{E^+}A-\tr_{E^-}A$; it extends $\Om^*(M)$-linearly
to $\str\colon\Om^*(M)\gtimes\Gam(\End(E))\to\Om^*(M)$.
Let $\nabla$ be a connection on $E$ preserving the grading.
$\nabla$ is an odd operator on $\Om^*(M,E)$.
If $L\in\Gam(\End(E)^-)$ is odd, then $D=\nabla+L$ is called a
{\em superconnection} on $E$; the {\em curvature}
$D^2=R+\nabla L+L^2\in(\Om^*(M)\otimes\Gam(\End(E)))^+$ is even.
With the superconnection, the Chern character of the virtual vector bundle
$E^+\ominus E^-$ can be represented by
\be\label{ch-super}
\ch_{\nabla\!,L}(E^+,E^-)=\str\;e^{\frac{\ii}{2\pi}D^2}.
\ee
It is a closed form on $M$ and its de Rham cohomology class is 
independent of the choice of $\nabla$ or $L$.
If $L$ is invertible everywhere on $M$ and the eigenvalues of $\ii L^2$ 
are negative, then (\ref{ch-super}) is exact:
\[ \ch_{\nabla\!,L}(E^+,E^-)=-{\textstyle \frac{\ii}{2\pi}}
\;d\!\int_1^\infty\!\!\!\str\,(e^{\frac{\ii}{2\pi}(\nabla+\beta L)^2}\!L)
\,d\beta. \]

Now let $E$ be an oriented real vector bundle of rank $r=2m$ over $M$
with a Euclidean structure $(\cdot,\cdot)$.
Suppose further that $E$ has a spin structure.
The associated spinor bundle $S(E)=S^+(E)\oplus S^-(E)$ is a graded 
complex vector bundle over $M$.
For any section $s\in\Gam(E)$, let $c(s)\in\Gam(\End(E)^-)$ be the
Clifford multiplication on $E$.
Then for any $s,s'\in\Gam(E)$, we have $\{c(s),c(s')\}=-2(s,s')$.
Given a connection $\nabla$ on $E$ preserving $(\cdot,\cdot)$, the induced
spinor connection $\nabla^S$ on $S(E)$ preserves the grading.
If $R$ is the curvature of $\nabla$, that of $\nabla^S$ is
$R^S=-\frac{1}{4}(\bfg,R\bfg)$, where $\bfg$ is now a section of
$E\otimes\cl(E)$.
For any $s\in\Gam(E)$, consider the superconnection
$D_s=\nabla^S+\left(\frac{\pi}{\ii}\right)^{\frac{1}{2}}c(s)$.
The Chern character form (\ref{ch-super}) of $S^+(E)\ominus S^-(E)$ is, 
using (\ref{supertrace}), 
\be\label{ch-spin}
\ch_{\nabla\!,s}(S^+(E),S^-(E))=(-1)^m
\hat A\left({\textstyle \frac{R}{2\pi}}\right)^{-\frac{1}{2}}
e_{\nabla\!,s}(E), 
\ee
where $e_{\nabla\!,s}(E)$ is given by (\ref{mq-euler}). 
In cohomology groups, (\ref{ch-spin}) reduces to
\[ \ch(S^+(E))-\ch(S^-(E))=(-1)^m\hat A(E)^{-\frac{1}{2}}\,e(E). \]

If $M$ is non-compact and the norm of $s$ increases rapidly away from 
$s^{-1}(0)$, then both sides of (\ref{ch-spin}) are differential forms
that decay rapidly away from $s^{-1}(0)$ and can represent cohomology
classes of such.
As before, we take the pull-back $\pi^*E$ with the tautological section
$\bfx$.
Then (\ref{ch-spin}) becomes
\be\label{ch-thom}
\ch_\nabla(\pi^*S^+(E),\pi^*S^-(E))=(-1)^m\pi^*\hat A
\left({\textstyle \frac{R}{2\pi}}\right)^{-\frac{1}{2}}\tau_\nabla(E), 
\ee
where $\tau_\nabla(E)$ is given by (\ref{mq-thom}).
Both sides of (\ref{ch-thom}) are forms on $E$ that decays exponentially 
in the fiber directions, hence it descends to an equality in $H^*(E,E\m M)$.
In the relative K-group $K(E,E\m M)$, the pair $\pi^*S^\pm(E)$
with the isomorphism $c(\bfx)$ away form the zero section is, up to a
factor of $(-1)^m$, the K-theoretic Thom class $i_!1\in K(E,E\m M)$.
Therefore (\ref{ch-thom}) reduces to the well-known formula
\[ \ch(i_!1)=\pi^*\hat A(E)^{-\frac{1}{2}}\,i_*1 \]
in cohomology groups $H^*(E,E\m M)$. 
The refinement (\ref{ch-thom}) as an equality of differential forms
is due to Mathai and Quillen (1986).
In fact, this is how (\ref{mq-thom}) was derived originally.

\sect{Equivariant Cohomology and Equivariant Vector Bundles}

\subsect{Equivariant cohomology}

Let $G$ be a compact Lie group with Lie algebra $\g$.
Fixing a basis $\{e_a\}$ of $\g$, the structure constants are given by
$[e_a,e_b]=t^c_{ab}\,e_c$.
Let $\{\vth^a\}$ and $\{\vph^a\}$ be the dual bases of $\g^*$ generating
the exterior algebra $\medwedge(\g^*)$ and the symmetric algebra $S(\g^*)$,
respectively.
The {\em Weil algebra} is $W(\g)=\medwedge(\g^*)\gtimes S(\g^*)$.
We define a grading on $W(\g)$ by specifying $\deg\vth^a=1$, $\deg\vph^a=2$.
The contraction $\iota_a$ and the exterior derivative $d$ are two
odd derivations on $W(\g)$ defined by
\be\label{weil}
\begin{array}{rl}
\iota_a\vth^b=\del^b_a,& \iota_a\vph^b=0, \\
d\,\vth^a=-\hf t^a_{bc}\,\vth^b\vth^c+\vph^a,&
d\,\vph^a=-t^a_{bc}\,\vth^b\vph^c.
\end{array}
\ee 
The Lie derivative is $L_a=\{\iota_a,d\}$.
These operators satisfy the usual (anti-)commutation relations 
\bea
d^2\eq0,\;\; L_a=\{\iota_a,d\},\;\; [L_a,d]=0, \label{comm0} \\
\!\!\!\!\!\!\{\iota_a,\iota_b\}\eq0,\;
[L_a,\iota_b]=t^c_{ab}\iota_c,\; [L_a,L_b]=t^c_{ab}L_c. \label{comm1} 
\eea
The cohomology of $(W(\g),d)$ is trivial.

If $G$ acts smoothly on a manifold $M$ on the left, let $V_a$ be the
vector field generated by the Lie algebra element $-e_a\in\g$.
Then $[V_a,V_b]=t^c_{ab}V_c$.
Denote $\iota_a=\iota_{V_a}$ and $L_a=L_{V_a}$, acting on $\Om^*(M)$.
In the {\em Weil model} of equivariant cohomology, one considers the
graded tensor product $W(\g)\gtimes\Om^*(M)$, on which the operators 
\bea
\tilde\iota_a\eq\iota_a\gtimes 1+1\gtimes\iota_a,    \nno
\tilde d\eq d\gtimes 1+1\gtimes d,   \nno 
\tilde L_a\eq L_a\gtimes 1+1\gtimes L_a  \nonumber
\eea
act and satisfy the same relations (\ref{comm0}) and (\ref{comm1}).
An element $\om\in W(\g)\gtimes\Om^*(M)$ is {\em basic} if it satisfies
$\iota_a\om=0$, $L_a\om=0$ for all indices $a$.
Let $\Om^*_G(M)=(W(\g)\gtimes\Om^*(M))_{\mathrm{bas}}$ be the set of such.
Elements of $\Om^*_G(M)$ are {\em equivariant differential forms} on $M$. 
The operator $\tilde d$ preserves $\Om^*_G(M)$ and its cohomology groups
$H^*_G(M)$ are the {\em equivariant cohomology groups} of $M$. 
They are isomorphic to the singular cohomology groups of $EG\times_GM$
with real coefficients.

The BRST model of Kalkman (1993) is obtained by applying an isomorphism
$\sig=e^{\vth^a\otimes\iota_a}$ of $W(\g)\gtimes\Om^*(M)$.
The operators become 
\bea
\sig\circ\tilde\iota_a\circ\sig^{-1}\eq\iota_a\gtimes 1,   \nno
\sig\circ\tilde d\circ\sig^{-1}
\eq \tilde d-\vph^a\gtimes\iota_a+\vth^a\gtimes L_a,   \nno 
\sig\circ\tilde L_a\circ\sig^{-1}\eq\tilde L_a.        \nonumber 
\eea
The subspace of basic forms in the Weil model becomes
$$\sig\,(\Om^*_G(M))=(S(\g^*)\otimes\Om^*(M))^G.$$
This is precisely the {\em Cartan model} of equivariant cohomology,
in which the exterior differential is
$$\tilde d'=1\otimes d-\vph^a\otimes\iota_a.$$

If $P$ is a principal $G$-bundle over a base space $B$, we can form 
an associated bundle $P\times_GM\to B$. 
Choose a connection on $P$ and let $\Th=\Th^ae_a\in\Om^1(P)\otimes\g$,
$\Ph=\Ph^ae_a\in\Om^2(P)\otimes\g$ be the connection, curvature forms,
respectively.
The components $\Th^a$, $\Ph^a$ satisfy the same relations (\ref{weil}).
Replacing $\vth^a$, $\vph^a$ by $\Th^a$, $\Ph^a$, we have a homomorphism
that maps $\om\in W(\g)\otimes\Om^*(M)$ to $\hat\om\in\Om^*(P\times M)$. 
If $\om$ is basic, then so is $\hat\om$, and the latter descends to 
a form $\bar\om$ on $P\times_GM$.
Furthermore, the operator $\tilde d$ on $\Om^*_G(M)$ descends to $d$
on $\Om^*(P\times_GM)$.
Thus we get the {\em Chern-Weil homomorphisms} 
$\Om^*_G(M)\to\Om^*(P\times_GM)$ and $H^*_G(M)\to H^*(P\times_GM)$.
For example, the vector space $\RE^r$ has an obvious $SO(r)$ action.
The Gaussian $r$-form (\ref{gaussian}) is invariant under $SO(r)$ and
can be extended to an $SO(r)$-equivariant closed $r$-form, called the
{\em universal Thom form}.
Let $E$ be an orientable real vector bundle $E$ of rank $r$ with a Euclidean
structure.
$E$ determines a principal $SO(r)$-bundle $P$; the associated bundle 
$P\times_{SO(r)}\RE^r$ is $E$ itself.
By applying the Chern-Weil homomorphism to this setting, we get a closed
$r$-form on $E$.
This is another construction of the Thom form (\ref{mq-thom}) 
by Mathai and Quillen (1986). 
Further information of equivariant cohomology can be found there and
in Guillemin (1999).

\subsect{Equivariant vector bundles}

Recall that a connection on a vector bundle $E\to M$ determines, 
for any $k\ge0$, a differential operator
$$\nabla\colon\Om^k(M,E)\to\Om^{k+1}(M,E).$$
The curvature $R=\nabla^2\in\Om^2(M,\End(E))$ satisfies the Bianchi
identity $\nabla R=0$.
If the connection preserves a Euclidean structure on $E$,
then $R$ is skew-symmetric.

If a Lie group $G$ acts on $M$ and the action can be lifted to $E$, then
$G$ also acts on the spaces $\Gam(E)$ and $\Om^*(M,E)$.
As before, the Lie derivatives $L_a$ on these spaces are the
infinitesimal actions of $-e_a\in\g$.
We choose a $G$-invariant connection on $E$.
The {\em moment} of the connection $\nabla$ under the $G$-action is
$\mu_a=L_a-\nabla_{V_a}$ acting on $\Gam(E)$.
In fact, $\mu_a$ is a section of $\End(E)$, 
or $\mu\in\Gam(\End(E))\otimes\g^*$.  
If a Euclidean structure on $E$ is preserved by both the connection
and the $G$-action, then $\mu_a$ is skew-symmetric.
On $\Om^*(M,E)$, we have
$$L_a=\{\iota_a,\nabla\}+\mu_a.$$

On the graded tensor product $W(\g)\gtimes\Om^*(M,E)$, the contraction 
$\tilde\iota_a$ and the Lie derivative $\tilde L_a$ act and satisfy
(\ref{comm1}).
In the Weil model, equivariant differential forms on $M$ with values in $E$
are the {\em basic} elements in $W(\g)\gtimes\Om^*(M,E)$, which form a 
subspace $\Om^*_G(M,E)=(W(\g)\gtimes\Om^*(M,E))_{\mathrm{bas}}$.
The {\em equivariant covariant derivative} is
\be\label{equiconn}
\tilde\nabla=d\gtimes 1+1\gtimes\nabla+\vth^a\gtimes\mu_a.
\ee
One checks that $\{\tilde\iota_a,\tilde\nabla\}=\tilde L_a$ and hence 
$\tilde\nabla$ preserves the basic subspace $\Om^*_G(M,E)$.
The {\em equivariant curvature} $\tilde R=\tilde\nabla^2$ is
\be\label{equicurv}
\tilde R=R-\vth^a\nabla\mu_a+\vph^a\mu_a+\hf\,\vth^a\vth^bR_{ab},
\ee
where $R_{ab}=R(V_a,V_b)\in\Gam(\End(E))$.
It satisfies the equivariant Bianchi identity $\tilde\nabla\tilde R=0$. 
Equivariant characteristic forms are invariant polynomials of $\tilde R$.
They are equivariantly closed and their equivariant cohomology classes 
do not depend on the choice of the $G$-invariant connection.
Hence they represent the {\em equivariant characteristic classes} of $E$ 
in $H^*_G(M)$.

For the BRST model, we use a similar isomorphism 
$\sig=e^{\vth^a\otimes\iota_a}$ on $W(\g)\gtimes\Om^*(M,E)$.
The operators become
\bea
\sig\circ\tilde\iota_a\circ\sig^{-1}\eq\iota_a\gtimes 1,   \nno
\sig\circ\tilde\nabla\circ\sig^{-1}
\eq\tilde\nabla-\vph^a\gtimes\iota_a+\vth^a\gtimes L_a,   \nno 
\sig\circ\tilde L_a\circ\sig^{-1}\eq\tilde L_a            \nonumber
\eea
and the basic subspace turns into
$$\sig\,(\Om^*_G(M,E))=(S(\g^*)\otimes\Om^*(M,E))^G.$$
This is the {\em Cartan model}, which can be found in Berline, Getzler
and Vergne (1992).
The equivariant connection is
$$\tilde\nabla'=1\otimes\nabla-\vph^a\otimes\iota_a.$$
The equivariant curvature is $\tilde R'=(\tilde\nabla')^2=R+\vph^a\mu_a$
and the characteristic forms are defined similarly.

Let $P\to B$ be a principal $G$-bundle with a connection $\Th$.
Following (\ref{equiconn}), the bundle $P\times E\to P\times M$ has 
a connection
$$\hat\nabla=d\otimes1+1\otimes\nabla+\Th^a\otimes\mu_a.$$
It descends to a connection $\bar\nabla$ on the vector bundle 
$P\times_GE\to P\times_GM$.
The map $\tilde\nabla\mapsto\bar\nabla$ can be considered as the analog of
the Chern-Weil homomorphism for connections.
There is also a homomorphism $\Om^*_G(M,E)\to\Om^*(P\times_GM,P\times_GE)$,
which commutes with the covariant derivatives $\tilde\nabla$, $\bar\nabla$.
The curvature $\bar R=\bar\nabla^2$ is the image of the equivariant 
curvature $\tilde R$.
Consequently, the equivariant characteristic forms descends to those
of $P\times_GE\to P\times_GM$ by the usual Chern-Weil homomorphism.

Now let $E=E^+\oplus E^-$ be a graded vector bundle over $M$ with 
a $G$-action preserving all the structures.
We have the $\Om^*_G(M)$-linear supertrace map 
$\str\colon\Om^*_G(M)\gtimes\Gam(\End(E))\to\Om^*_G(M)$.
If $\nabla$ is a $G$-invariant connection on $E$ preserving the grading 
and if $L\in\Gam(\End(E)^-)^G$ is odd and $G$-invariant, then
$\tilde D=\tilde\nabla+L$ is an {\em equivariant superconnection}.
The equivariant counterpart of (\ref{ch-super}) is
$$\ch_{\tilde\nabla\!,L}(E^+,E^-)=\str\;e^{\frac{\ii}{2\pi}\tilde D^2}
\in\Om^*_G(M),$$
representing the equivariant Chern character of $E^+\ominus E^-$ in 
$H^*_G(M)$.

\subsect{Representatives of the equivariant Euler and Thom classes}

Consider an oriented real vector bundle $E\to M$ of rank $r$ 
with a Euclidean structure $(\cdot,\cdot)$. 
Choose a connection $\nabla$ on $E$ preserving $(\cdot,\cdot)$.
We assume that a Lie group $G$ acts on $M$ and that the action can be
lifted to $E$ preserving all the structures on $E$.
We use the Weil model; the constructions in the Cartan model are similar.
For any $\al\in\Om_G^k(M,E)$ and $\beta\in\Om_G^l(M,E)$, we obtain
$(\al,\wedge\beta)\in\Om_G^{k+l}(M)$ by taking the wedge product of forms 
as well as the pairing in $E$.
The Berezin integral of $\om\in\Om_G^*(M,\medwedge^*E^*)$ along the fibers 
of $E$ is $\bint\om=\bra\nu,\om\ket\in\Om_G^*(M)$.  
Here $\nu$ is the unit section of the canonically trivial determinant 
line bundle $\medwedge^rE$, compatible with the orientation of $E$. 
The equivariant Euler form 
\be\label{equieuler}
e_{\tilde\nabla}(E)=\frac{1}{(2\pi)^{\frac{r}{2}}}
\bint e^{\frac{1}{2}(\,\cdot\,,\tilde R\,\cdot\,)}
=\Pf\left({\textstyle \frac{\tilde R}{2\pi}}\right)
\ee
is equivariantly closed.
It represents the equivariant Euler class $e_G(E)\in H^*_G(M)$.

Given a $G$-invariant section $s\in\Gam(E)^G$, the equivariant counterpart
of (\ref{mqexp}) is
\be
\SS_{\tilde\nabla\!,s}=\hf(s,s)+(\tilde\nabla s,\cdot\,)
       +\hf(\,\cdot\,,\tilde R\,\cdot\,)
\ee
and that of Mathai-Quillen's Euler form (\ref{mq-euler}) is
\be\label{mq-eq-euler}
e_{\tilde\nabla\!,s}(E)=\frac{(-1)^{\frac{r(r+1)}{2}}}{(2\pi)^{\frac{r}{2}}}
\bint e^{-\SS_{\tilde\nabla\!,s}}.
\ee
It is also equivariantly closed, and its equivariant cohomology class 
is $e_G(E)$.
The equivariant extension of Mathai-Quillen's Thom form (\ref{mq-thom}) is
\be
\tau_{\tilde\nabla}(E)=\frac{(-1)^{\frac{r(r+1)}{2}}}{(2\pi)^{\frac{r}{2}}} 
    \bint e^{-\frac{1}{2}(\sbfx,\sbfx)-(\tilde\nabla\sbfx,\cdot\,)
                  -\frac{1}{2}(\,\cdot\,,\tilde R\,\cdot\,)},
\ee
where $\bfx$ is the ($G$-invariant) tautological section of $\pi^*E\to E$.

Finally, $G$ acts on the (graded) spinor bundle $S(E)$. 
Using the equivariant superconnection 
$\tilde D_s=\tilde\nabla^S+\left(\frac{\pi}{\ii}\right)^{\frac{1}{2}}c(s)$,
(\ref{ch-spin}) generalizes to
$$
\ch_{\tilde\nabla\!,s}(S^+(E),S^-(E))=(-1)^m
\hat A{\textstyle \left(\frac{\tilde R}{2\pi}\right)}^{-\frac{1}{2}}
e_{\tilde\nabla\!,s}(E). 
$$
Now apply the construction to the bundle $\pi^*E\to E$ and its 
tautological section $\bfx$.
The pair $\pi^*S^\pm(E)$ with an odd bundle map $c(\bfx)$ determines,
up to a factor of $(-1)^m$, the Thom class $i_!1_G$ in the equivariant
K-group $K_G(E,E\m M)$.
The equivariant analog of (\ref{ch-thom}) descends to
$$\ch_G(i_!1_G)=\pi^*\hat A_G(E)^{-\frac{1}{2}}i_*1_G$$
in equivariant cohomology.

\sect{Superspace Formulation}

\subsect{Mathai-Quillen formalism and the superspace $\RF$} 

Let $\RF$ be the superspace with one fermionic coordinate $\th$ but
no bosonic coordinates.
The translation on $\RF$ is generated by $D=\frac{\pdr}{\pdr\th}$,
which satisfies $\{D,D\}=0$.
We consider a sigma model on $\RF$ whose target space is an (ordinary)
smooth manifold $M$ of dimension $n$.
A map $X\colon\RF\to M$ can be written as $X(\th)=x+\ii\th\psi$.
Here $x=X\bos\in M$ and $\psi=-\ii DX\bos\in T_xM$; the latter is fermionic.
Under the translation $\th\mapsto\th+\eps$, $x$ and $\psi$ vary
according to the {\em supersymmetry transformations}
\be\label{susy-map}
\begin{array}{l}
\del x=\eps DX\bos=\ii\eps\psi,\\
\del\psi=\eps D(DX)\bos=0. 
\end{array}
\ee
Clearly, $\del^2=0$, which is also a consequence of $D^2=0$.

For any $p$-form $\om\in\Om^p(M)$, we have an {\em observable}
$\OO_\om(X)=\frac{1}{p!}X^*\om(D,\cdots,D)\bos$.
In local coordinates, 
$\om=\frac{1}{p!}\om_{i_1\cdots i_p}(x)\,dx^{i_1}\wedge\cdots\wedge dx^{i_p}$
and $\OO_\om(x,\psi)=
\frac{\ii^{\,p}}{p!}\om_{i_1\cdots i_p}(x)\psi^{i_1}\cdots\psi^{i_p}$.
Using $C(\cdot)$ to denote the set of function(al)s on a space, we can
identify $C(\Map(\RF,M))$ with $\Om^*(M)$.
Under (\ref{susy-map}), $\del\OO_\om(X)=\eps\,\OO_{d\om}(X)$.
So $\OO_\om(X)$ is invariant under supersymmetry if and only if $\om$
is closed.
The cohomology of $\del$ is the de Rham cohomology of $M$.
Consider the measure $[dX]=[dx][d\psi]$.
In local coordinates, $[dx]=dx^1\cdots dx^n$ is the standard (bosonic) measure
and $[d\psi]=d\psi^1\cdots d\psi^n$ is a fermionic
measure such that $\int[d\psi](-1)^{\frac{n(n-1)}{2}}\psi^1\cdots\psi^n=1$.
For any $\om\in\Om^n(M)$, the superfield integral $\int[dX]\,\OO_\om(X)$ is
equal to the usual integral $\int_M\om$ if the latter exists.

Let $E\to M$ be a real vector bundle of rank $r$ with an inner product 
$(\cdot,\cdot)$, and let $\nabla$ be a compatible connection whose curvature
is $R$.
Consider a theory whose fields are $X\in\Map(\RF,M)$ and a fermionic section 
$\XI\in\Gam(X^*E)$.
Let $\DD=(X^*\nabla)_D$ be the covariant derivative along $D$ in the 
pull-back bundle $X^*E\to\RF$.
Then $\chi=\XI\bos\in E_x$ is fermionic and $f=\DD\XI\bos\in E_x$ is bosonic.

Given a fixed section $s\in\Gam(E)$, we write a superspace action
\bea
\vc S_\MQ[X,\XI]=\int_{\RF}d\th\,(\XI,\hf\DD\XI+\ii s\circ X)  \label{S-MQ}\\
\eq \hf(f,f)+\ii(f,s)-(\nabla_\psi s,\chi)+\qt(\chi,R(\psi,\psi)\chi).\nonumber
\eea
It is automatically supersymmetric.
Performing the Gaussian integral over $f$ and replacing $\chi$ by $-\ii\chi$,
we get
\be\label{susy-mq}
\int[d\XI]\,e^{-S_\MQ[X,\XI]}=\frac{1}{(2\pi)^{\frac{r}{2}}}\int[d\chi]
\,e^{-S_\MQ[x,\psi,\chi]},
\ee
where
\be\label{susy-action}
S_\MQ[x,\psi,\chi]=\hf(s,s)-\ii(\chi,\nabla_\psi s)-\qt(\chi,R(\psi,\psi)\chi).
\ee
Actually, (\ref{susy-mq}) holds up to possible factors of $\ii$ which we
ignore here and in subsequent partition functions and expectations values.
With this caveat, (\ref{susy-mq}) is equal to $\OO_{e(\nabla\!,s)(E)}(X)$,
where $e(\nabla\!,s)(E)$ is given by (\ref{mq-euler}).
Furthermore, for any closed form $\om$ on $M$, the expectation value
\be\label{mq-exp}
\bra\OO_\om(X)\ket=\int[dX][d\XI]\,\OO_\om(X)\,e^{-S_\MQ[X,\XI]}
\ee
is equal to (\ref{mq-pd}).

\subsect{Equivariant cohomology and gauged sigma model on $\RF$}

Suppose $G$ is a Lie group and $P$ is a principal $G$-bundle over $\RF$.
Since $\th$ is nilpotent, we can choose a ``trivialisation'' of $P$
such that the connection and curvature are $A\in\Om^1(\RF)\otimes\g$ and
$F\in\Om^2(\RF)\otimes\g$, respectively.
($\g$ is the Lie algebra of $G$.)
In components, $c=\ii\,\iota_DA\in\g$ is fermionic and
$\phi=-\frac{\ii}{2}\iota_D^2F\in\g$ is bosonic.
The space of connections $\A$ is the set of pairs $(c,\phi)$.
Under $\th\mapsto\th+\eps$,
\be\label{susy-A}
\begin{array}{l}
\del c=\eps(\phi+\hfi[c,c]),\\
\del\phi=\ii\eps[c,\phi].
\end{array}
\ee
Thus the algebra $C(\A)$ is isomorphic to the Weil algebra $W(\g)$ and 
$\del$ corresponds to the differential $d$ in (\ref{weil}).
This relation between gauge theory on a fermionic space and the Weil
algebra can be found in Blau and Thompson (1997). 

With a trivialisation of $P$, the group of gauge transformation $\GG$
can be identified with $\Map(\RF,G)$.
Any group element is of the form $\gh=g\,e^{\ii\xi\th}$, with $g=\gh\bos\in G$
and $\xi=\ii\,\iota_D\gh^*\vpi\in\g$ (fermionic), where $\vpi$ is the 
Maurer-Cartan form on $G$.
The action of $\gh$ is $A\mapsto A'=\Ad_\gh(A-\gh^*\vpi)$, or
$c\mapsto c'=\Ad_g(c-\xi)$ and $\phi\mapsto\phi'=\Ad_g\phi$.
By choosing $\xi=c$, we obtained a new trivialisation, called the 
{\em Wess-Zumino gauge}, in which $c'=0$.
The residual gauge redundancy is $G$, and $\A/\GG=\g/\Ad_G$.
The Wess-Zumino gauge is not preserved by the translation on $\RF$ unless
we define $\del'$ by composing $\del$ with a suitable (infinitesimal)
gauge transformation.
If so, then $\del'\phi=0$.

Suppose $M$ is a manifold with a left $G$-action.
As before, let $\{e_a\}$ be a basis of $\g$ and let the vector field $V_a$
be the infinitesimal action of $-e_a$.  
In the gauged sigma model, we include another field $X\in\Gam(P\times_GM)$.
With a trivialisation of $P$, we can identify $X$ with a map $X\colon\RF\to M$.
The covariant derivative is given by $\nabla X=dX-A^aV_a$, $\DD X=\nabla_DX$.
Let $x=X\bos\in M$ and $\psi=-\ii\,\DD X\bos\in T_xM$.
Then the supersymmetric transformations are
\be\label{susy-gmap}
\begin{array}{l}
\del x^i=\ii\,\eps(\psi^i-c^aV^i_a),\\
\del\psi^i=-\eps(\phi^aV_a^i+\ii\,c^jV^i_{a,j}).
\end{array}
\ee
In the Wess-Zumino gauge, the transformations simplify to
$\del' x=\ii\eps\psi$, $\del'\psi=-\eps\phi^aV_a$.

The observables form the $\GG$-invariant part of the space
$C(\A\times\Map(\RF,M))$.
For any $\om\in\Om^p(M)$, we have
\bea
\!\!\!\!\!\!\OO_\om(X, A)\eq{\textstyle \frac{1}{p!}}
                   \om(\DD X,\cdots,\DD X)\bos                         \nno
                \eq{\textstyle \frac{\ii^{\,p}}{p!}}
		   \om_{i_1\cdots i_p}(x)\psi^{i_1}\cdots\psi^{i_p}.
\eea
$\OO_\om(X,A)$ is gauge covariant: $\OO_\om(X,A)\mapsto\OO_{g^*\om}(X,A)$,
and the set of gauge invariant observables is thus identified with 
$(S(\g^*)\times\Om^*(M))^G$.
Moreover, since
\bea
\del\OO_\om(X,A)\eq\eps(\OO_{d\om}(X,A)-\ii c^a\OO_{L_a\om}(X,A)\nno
                \vc-\ii\phi^a\OO_{\iota_a\om}(X,A)),  \nonumber
\eea
$\del$ corresponds to the differential $\tilde d'$ in BRST model.

Let $E\to M$ be an equivariant vector bundle and let $\nabla$ be a 
$G$-invariant connection with curvature $R$ and moment $\mu$.
Any $s\in\Gam(E)^G$ defines a section of $P\times_GE\to P\times_GM$, 
still denoted by $s$.
Consider a theory with superfields $X\in\Gam(P\times_GM)$ and 
$\XI\in\Gam(X^*(P\times_GE))$ (fermionic).
Let $\DD$ be the covariant derivative of the pull-back connection. 
With a trivialisation of $P$, we put $\chi=\XI\bos\in E_x$ (fermionic)
and $f=\DD\XI\bos\in E_x$ (bosonic).
The equivariant extension of (\ref{S-MQ}) is 
\[ S_\MQ[X,\XI,A]=\int_\RF d\th\,(\XI,\hf\DD\XI+\ii s\circ X)    \]
Similar to (\ref{susy-mq}), we get, in the Wess-Zumino gauge,
\be\label{susy-eq-mq}
\int[d\XI]\,e^{-S_\MQ[X,\XI,A]}=\frac{1}{(2\pi)^{\frac{r}{2}}}\int[d\chi]
\,e^{-S_\MQ[x,\psi,\phi,\chi]},
\ee
where
\bea
\vc S_\MQ[x,\psi,\phi,\chi]=\hf(s,s)-\ii(\chi,\nabla_\psi s)     \nno
\vc\quad\quad-\qt(\chi,R(\psi,\psi)\chi)-\hfi(\chi,\phi^a\mu_a\chi).
\eea
(\ref{susy-eq-mq}) is equal to $\OO_{\tilde e(\nabla\!,s)}(X,A)$, where
$\tilde e(\nabla\!,s)$ is given by (\ref{mq-eq-euler}).

\subsect{The Atiyah-Jeffrey formula}

Given the $G$-action on $M$, for any $x\in M$, there is a linear map 
$C_x\colon\g\to T_xM$ defined by $C_x(e_a)=V_a(x)$.
With an invariant inner product $(\cdot,\cdot)$ on $\g$ and an invariant
Riemannian metric on $M$, the adjoint of $C_x$ is 
$C_x^\dagger\colon T_xM\to\g$, that is, $C^\dagger\in\Om^1(M)\otimes\g$.
If $G$ acts on $M$ freely, then $C_x$ is injective and $(C^\dagger C)_x$
is invertible for all $x\in M$.  
The projection $M\to\bar M=M/G$ is a principal $G$-bundle.
It has a connection such that the horizontal subspace is the orthogonal
compliment of the $G$-orbits.
The connection $1$-form is $\Th=(C^\dagger C)^{-1}C^\dagger$ whereas the
curvature is $\Ph=(C^\dagger C)^{-1}dC^\dagger$ on horizontal vectors.  

Let $\om$ be an equivariant form on $M$.
Suppose $G$ acts on $M$ freely, then $\om$ descends to a form $\bar\om$ 
on $\bar M$.
We look for a gauge invariant, supersymmetric quantity $\Ups(X,A)$
such that
\be\label{proj}
\frac{1}{\vol(\GG)}\int[dX][dA]\,\OO_\om(X,A)\Ups(X,A)
=\int[d\bar X]\,\OO_{\bar\om}(\bar X).
\ee
Mathematically, $\Ups$ corresponds to a closed equivariant form $\ups$ on $M$
such that
$$
\frac{1}{\vol(G)}\int_{\phi\in\g}[d\phi]\int_M\om(\phi)\wedge\ups(\phi)
=\int_{\bar M}\bar\om,
$$
which is (\ref{proj}) in the Wess-Zumino gauge.
$\ups$ can be understood as an equivariant homology cycle, as in
Austin and Braam (1995).

Let $P$ be a $G$-bundle over $\RF$ with a connection and let
$\Ad P=P\times_G\g\to\RF$ be the adjoint bundle.
Consider a (bosonic) superfield $\Lam\in\Gam(\Ad P)$.
Put $\lam=\Lam\bos$ (bosonic) and $\eta=-\ii\DD\Lam\bos$ (fermionic).
Choosing a trivialisation of $P$, $\lam$ and $\eta$ are both in $\g$.
Under $\th\mapsto\th+\eps$, they transform as
\be\label{susy-lam}
\begin{array}{l}
\del\lam=\ii\eps(\eta+[c,\lam]), \\
\del\eta=\eps([\phi,\lam]-\ii[c,\eta]).
\end{array}
\ee
The superspace action 
\[ S_\CMR[X,\Lam,A]=\ii\int_\RF d\th\,(\Lam,C^\dagger\DD X)  \]
is invariant under (\ref{susy-A}), (\ref{susy-gmap}) and (\ref{susy-lam})
and, under the Wess-Zumino gauge, it is
\bea\label{S-CMR}
\vc S_\CMR[x,\psi,\phi,\eta,\lam]=-\ii(\eta,C^\dagger\psi)         \nno
\vc\quad-\ii(\lam,dC^\dagger(\psi,\psi))+(\lam,C^\dagger C\phi).
\eea
If $G$ acts on $M$ freely, then
\be
\Ups(X,A)=\int[d\Lam]\,e^{-S_\CMR[X,\Lam,A]}
\ee
satisfies (\ref{proj}).
The factor $\Ups(X,A)$ in (\ref{proj}) is called {\em projection} in
Cordes, Moore and Ramgoolam (1996).

Let $E\to M$ be a $G$-equivariant vector bundle with a fixed $G$-invariant
connection $\nabla$, moment $\mu$, and an invariant section $s$.
Consider the superspace action
\[ S_\AJ[X,\XI,\Lam,A]=S_\MQ[X,\XI,A]+S_\CMR[X,\Lam,A]. \]
In the Wess-Zumino gauge and after the Gaussian integral over $f$,
it becomes the Atiyah-Jeffrey action
\be\label{S-AJ}
S_\AJ[x,\psi,\phi,\chi,\eta,\lam]
=S_\MQ[x,\psi,\phi,\chi]+S_\CMR[x,\psi,\phi,\eta,\lam].
\ee
If $s$ intersect the zero section transversely and $G$ acts on $s^{-1}(0)$
freely, then $s^{-1}(0)/G$ is smooth and
\bea\label{AJ}
\vc\int_{s^{-1}(0)/G}\bar\om=\int[dx][d\psi][d\phi][d\chi][d\eta][d\lam] \nno
\vc\quad\quad\OO_\om(x,\psi,\phi)\,e^{-S_\AJ[x,\phi,\phi,\chi,\eta,\lam]}.
\eea
for any closed equivariant form $\om$ on $M$.
(\ref{AJ}) is the formula of Atiyah and Jeffrey (1990) and of Witten (1988a)
in an infinite dimensional setting.
When $s^{-1}(0)/G$ is not smooth, the right-hand side of (\ref{AJ}) 
can be regarded as a definition of the left-hand side.

It is often convenient to add to $S_\AJ$ another term
\bea\label{S'}
\vc\Del S[X,\Lam,A]=-\frac{1}{4}\int_\RF([\iota_D^2F,\Lam],\DD\Lam) \nno
\vc\quad=\hfi(\phi,[\eta,\eta])+\hf([\phi,\lam],[\phi,\lam]).
\eea
Since (\ref{S'}) is $\del$-exact and no new field is added, the integral
(\ref{AJ}) does not change if $\Del S$ is added to $S_\AJ$.

\sect{Applications to Cohomological Field Theories}

We now apply the Mathai-Quillen construction formally to a number of cases
in which the both the rank of the vector bundle and the dimension of the 
base space are infinite.
Thus the (bosonic and fermionic) integrals in (\ref{mq-exp}) or (\ref{AJ})
become path integrals in quantum mechanics or quantum field theory. 

\subsect{Supersymmetric quantum mechanics}

Let $(M, g)$ be a Riemannian manifold and $LM=\Map(S^1,M)$, the loop space.
At each point $u\in LM$, which is a map $u\colon S^1\to M$, the tangent
space is $T_uLM=\Gam(u^*TM)$.
In particular, $\dot u=\frac{du}{dt}$, where $t$ is a parameter on $S^1$,
is a tangent vector at $u$ and $u\mapsto\dot u$ is a vector field on $LM$.
For any Morse function $h$ on $M$, $\s(u)=\dot u+(\grad h)\circ u$ is
another vector field on $LM$.

Vector fields on $LM$ can be identified as sections of the bundle 
$\ev^*TM\to S^1\times LM$, where $\ev\colon S^1\times LM\to M$ is the 
evaluation map.
The Levi-Civita connection $\nabla$ on $TM$ pulls back to a connection on 
$\ev^*TM$ and the covariant derivatives along $LM$ define a natural connection 
$\nabla^{LM}$ on $T(LM)$.
For example,  for any tangent vector $V\in T_uLM=\Gam(u^*TM)$, we have
$\nabla^{LM}_V\s(u)=\nabla^u_t V+(\nabla_V\grad h)\circ u$, where 
$\nabla^u_t$ is the pull-back connection on $u^*TM$.
The Riemann curvature tensor $R$ on $M$ determines that on $LM$.

The (infinite dimensional) analog of (\ref{susy-mq}) is
\be\label{par-qm}
\int [du][d\psi][d\chi]\;e^{-\int_{S^1}dt\,L[u,\psi,\chi]},
\ee
where $\psi,\chi\in T_uLM=\Gam(u^*TM)$ are fermionic and 
\bea\label{lag-qm}
\quad L[u,\psi,\chi]\eq\hf g(\dot u+\grad h,\dot u+\grad h)   \nno
\vc-\ii g(\chi,\nabla^u_t\psi+\nabla_\psi\grad h)              \nno
\vc-\qt g(\chi,R(\psi,\psi)\chi).
\eea
(\ref{lag-qm}) is, up to a total derivative, the Lagrangian of the Euclidean
$N=2$ supersymmetric quantum mechanics on $M$.
The partition function (\ref{par-qm}) is equal to Euler characteristic number
of $LM$ or $M$, which can be confirmed by an (exact) stationary phase 
calculation.

\subsect{Topological sigma model}

Let $\Sig$ be a Riemann surface $\Sig$ with complex structure $\veps$ and
let $(M,\om)$ be a symplectic manifold with a compatible almost complex
structure $J$.  
Let $\EE$ be a vector bundle over $\Map(\Sig,M)$ so that the fiber over $u$
is $\EE_u=\Gam(u^*TM\times T^*\Sig)$.
For any $u\in\Map(\Sig,M)$, $du\in\EE_u$ and $u\mapsto du$ is a section of
$\EE$.
The pull-back of the Levi-Civita connection on $TM$, tensored with
a connection on $T^*\Sig$, defines a connection on $\EE$.

The vector bundle to which we apply the Mathai-Quillen formalism is the 
anti-holomorphic part $\EE^{01}$ of $\EE$.
The fiber over $u\in\Map(\Sig,M)$ is 
$\EE^{01}_u=\Gam((u^*TM\otimes T^*\Sig)^{01})$.
The sub-bundle $\EE^{01}$ has a connection $\nabla^{01}$ via projection
from $\EE$. 
$\EE^{01}$ has a natural section 
$\s\colon u\mapsto\bar\pdr_Ju=\hf(du+J\circ du\circ\veps)$.
Solutions to the equation $\bar\pdr_Ju=0$ are {\em pseudo-holomorphic} (or
{\em $J$-holomorphic}) curves; let $\MM=\s^{-1}(0)$ be the space of such.
Its (virtual) dimension is
\be
\dim\MM=\hf\chi(\Sig)\dim M+2c_1(u^*TM).
\ee
Along any $V\in T_u\Map(\Sig,M)=\Gam(u^*TM)$, the covariant derivative of 
$\s=\bar\pdr_J$ is calculated in Wu (1995):
\be
\nabla^{01}_V(\bar\pdr_J)=\hf(\nabla^uV+J\circ\nabla^uV\circ\veps)+
\qt\nabla_VJ\circ(du\circ\veps+J\circ du),
\ee
where $\nabla^u$ is the pull-back connection on $u^*TM$.

To write the Mathai-Quillen formalism for the bundle $\EE^{01}\to\Map(\Sig,M)$,
we let $\psi\in\Gam(u^*TM)$ and $\chi\in\Gam((u^*TM\otimes T^*\Sig)^{01})$ be
fermionic fields.
(\ref{susy-action}) becomes the Lagrangian
\bea\label{lag-sig}
\vc L[u,\psi,\chi]=\hf||du||^2+\hf(du,J\circ du\circ\veps)      \nno
\vc\quad\quad-\ii(\chi,\nabla^u\psi+(\nabla_\psi J)\circ du\circ\veps)  \nno
\vc\quad\quad
-{\textstyle \frac{1}{8}}(\chi,(R(\psi,\psi)-\hf(\nabla_\psi J)^2)\chi).
\eea
It is precisely the Lagrangian of the topological sigma model of
Witten (1988b).
Here the pairing $(\cdot,\cdot)$ is induced by the Riemannian metric
$\om(\cdot,J\cdot)$ on $M$ and a metric on $\Sig$ that is compatible with
$\veps$.
The second term in (\ref{lag-sig}), integrated over $\Sig$, is equal to 
$\int_\Sig u^*\om=\bra[\om],u_*[\Sig]\ket$.

For any differential form $\al\in\Om^p(M)$, let $\OO_\al(u,\psi)$ be
the observable obtained from $\ev^*\al\in\Om^p(\Sig\times\Map(\Sig,M))$ 
by identifying $\Om^*(\Map(\Sig,M))$ with $C(\Map(\RF,\Map(\Sig,M)))$.
If $\al$ is closed and $\gam\in H_q(\Sig)$ is a homology cycle, then
$W_{\al,\gam}(u,\psi)=\int_\gam\OO_\al(u,\psi)$ is identified with a 
closed $(p-q)$-form on $\Map(\Sig,M)$.
For closed $\al_i\in\Om^{p_i}(M)$ and $\gam_i\in H_{q_i}(\Sig)$ 
($1\le i\le r$), the expectation values
\be\label{GM}
\left\bra\prod_{i=1}^r W_{\al_i,\gam_i}\!\right\ket=\int[du][d\psi][d\chi]
\prod_{i=1}^r W_{\al_i,\gam_i}(u,\psi)\;e^{-S[u,\psi,\chi]}
\ee
are the {\em Gromov-Witten invariants} of $(M,\om)$. 
Moreover, (\ref{GM}) is non-zero only if $\sum_{i=1}^r(p_i-q_i)=\dim\MM$.

\subsect{Topological gauge theory}

Let $M$ be a compact, oriented four-manifold, $G$, a compact, semisimple
Lie group, and $P\to M$, a principal $G$-bundle.
Denote by $\A$ the space of connections on $P$ and $\GG$, the group of
gauge transformations.
The Lie algebra of $\GG$ is $\Lie(\GG)=\Gam(\Ad P)=\Om^0(M,\Ad P)$.
At $A\in\A$, the tangent space is $T_A\A=\Om^1(M,\Ad P)$.
Both spaces have inner products if we choose an invariant inner product 
$(\cdot,\cdot)$ on the Lie algebra $\g$ of $G$ and a Riemannian
metric $g$ on $M$.
The infinitesimal action of $\GG$ on $\A$ is 
$C=\nabla_A\colon\Lie(\GG)\to T_A\A$.

With a Riemannian metric, any $2$-form on $M$ decomposes into self-dual
and anti-self-dual parts: $\Om^2(M)=\Om^2_+(M)\oplus\Om^2_-(M)$.
We consider a trivial vector bundle $\EE\to\A$ whose fiber is
$\Om^2_+(M,\Ad P)$.  
$\GG$ acts on $\EE$ and the bundle is $\GG$-equivariant.
The trivial connection on $\EE$ is $\GG$-invariant; the moment is given
by $\phi\in\Gam(\Ad P)\colon\chi\in\Om^2_+(M,\Ad P)\mapsto[\phi.\chi]$.
The bundle $\EE$ has a natural section $\s\colon A\in\A\mapsto F^+_A$, the 
self-dual part of the curvature.
Its derivative along $V\in\Om^1(M,\Ad P)=T_A\A$ is $L_V\s=(\nabla_AV)^+$.
The section $\s$ is $\GG$-invariant, the zero set $\s^{-1}(0)$ is the
space of anti-self-dual connections, and the quotient $\MM=\s^{-1}(0)/\GG$
is the {\em instanton moduli space}.
Its (virtual) dimension is
\[ \dim\MM=4\check h(\g)k(P)-\hf\dim G(\chi(M)+\sig(M)),
\]
where $\check h(\g)$ is the dual Coxeter number of $\g$ and 
$k(P)=-\frac{1}{4\check h(\g)}\bra p_1(\Ad P),[M]\ket\in\ZZ$ is the instanton
number of $P$. 

We proceed with the Mathai-Quillen interpretation of Atiyah and Jeffrey (1990).
Let $\psi\in\Om^1(M,\Ad P)$, $\chi\in\Om^2_+(M,\Ad P)$, $\eta\in\Gam(\Ad P)$
be fermionic fields and $\phi,\lam\in\Gam(\Ad P)$, bosonic fields.
The combination of (\ref{S-AJ}) and (\ref{S'}) is given by the Lagrangian
\bea\label{lag-tym}
\vc L[A,\psi,\phi,\chi,\eta,\lam]=
    \hf||F^+_A||^2+(\phi,\nabla_A^\dagger\nabla_A\lam)        \nno
\vc\quad -\ii(\eta,\nabla_A\psi)-\ii(\chi,\nabla_A\psi)
         -\ii(\lam,[\psi,\psi])                               \nno
\vc\quad +\hfi(\phi,[\chi,\chi]+[\eta,\eta])-\hf||[\phi,\lam]||^2.
\eea
With an additional topological term proportional to $(F_A,\wedge F_A)$,
(\ref{lag-tym}) is the Lagrangian of topological gauge theory of 
Witten (1988a).
Here $(\cdot,\cdot)$ is the pairing induced by a Riemannian metric on $M$
and an invariant inner product on $\g$. 

There is a tautological connection on the $G$-bundle 
$\A\times P\to\A\times M$. 
It is invariant under the $\GG$-action.
Identifying $\Om^*(\A)$ with $C(\Map(\RF,\A))$ and using the Cartan model,
the $\GG$-equivariant curvature is $\FF=F_A+\ii\psi+\phi$.
For any homology cycle $\gam\in H_q(M)$, 
\be
W_\gam(A,\psi,\phi)=\frac{1}{4\check h(\g)}\int_\gam(\FF,\wedge\FF)
\ee
corresponds to a closed $\GG$-equivariant form on $\A$.
For $\gam_i\in H_{q_i}(M)$ ($1\le i\le r$), the expectation values
\bea\label{D}
\vc\left\bra\prod_{i=1}^r W_{\gam_i}\right\ket=\frac{1}{\vol(\GG)}
\int[dA][d\psi][d\phi][d\chi][d\eta][d\lam]                 \nno
\vc\quad\quad\quad\prod_{i=1}^r W_{\gam_i}(A,\psi,\phi)
\,\;e^{-S[A,\psi,\phi,\chi,\eta,\lam]}
\eea
are, up to a factor of $|Z(G)|$, Donaldson invariants of $M$.
Moreover, (\ref{D}) is non-zero only if $\sum_{i=1}^r(4-q_i)=\dim\MM$.

Other cohomological field theories can also be understood or constructed
by the Mathai-Quillen formalism.
Of such we mention only the topological field theories of Abelian and 
non-Abelian monopoles in Labastida and Mari\~no (1995), which are related
to the Seiberg-Witten invariants.

\vspace{1cm}

\noindent
{\Large\bf See also}\\

\noindent
{\bf Topological quantum field theory: overview.
Donaldson-Witten theory.
Topological sigma models. 
Equivariant cohomology and the Weil and Cartan models.
Characteristic classes.}

\vspace{1cm}

\noindent
{\Large\bf Further Reading}\\

Atiyah, M F and Jeffrey, L C (1990) Topological Lagrangians and cohomology.
{\em J.\ Geom.\ Phys.} 7: 119-138.

Austin, D M and Braam, P J (1995) {\em Equivariant homology}.
{\em Math.\ Proc.\ Camb.\ Phil.\ Soc.} 118: 125-139.

Berline, N, Getzler, E and Vergne, M (1992) {\em Heat kernels and Dirac
operators}. Berlin: Springer-Verlag.

Blau, M and Thompson, G (1997) Aspects of $N_T\ge2$ topological gauge
theories and D-branes. {\em Nucl.\ Phys.\ B} 492: 545-590.
 
Cordes, S, Moore, M and Ramgoolam, S (1996) Lectures on $2$D Yang-Mills
theory, equivariant cohomology and topological field theories. 
(Les Houches, 1994), pp.\ 505-682, North-Holland, Amsterdam.

Guillemin, V W and Sternberg, S (1999) {\em Supersymmetry and equivariant
de Rham theory}. Berlin: Springer-Verlag.

Kalkman, J (1993) BRST model for equivariant cohomology and representatives
for the equivariant Thom class. {\em Commun.\ Math.\ Phys.} 153: 447-463.

Labastida, J M F and Mari\~no, M (1995) A topological lagrangian for monopoles
on four-manifolds. {\em Phys.\ Lett.\ B} 351: 146-152;
Non-abelian monopoles on four-manifolds. {\em Nucl.\ Phys.\ B} 448: 373-395. 

Mathai, V and Quillen, D (1986) Superconnections, Thom classes, and
equivariant differential forms. {\em Topology} 25: 85-100.

Witten, E (1988a) Topological quantum field theory. 
{\em Commun.\ Math.\ Phys.} 117: 353-386.

Witten, E (1988b) Topological sigma models. {\em Commun.\ Math.\ Phys.}
118: 411-449.

Wu, S (1995) On the Mathai-Quillen formalism of topological sigma models.
{\em J. Geom. Phys.} 17: 299-309.

Zhang, W (2001) {\em Lectures on Chern-Weil theory and Witten deformations},
Nankai Tracts in Mathematics, Vol.\ 4. Singapore: World Scientific.

\end{document}